\documentclass{article}
\usepackage{spconf,amsmath,graphicx,hyperref}
\usepackage{booktabs}   
\usepackage{tabularx}   
\usepackage{multirow}   
\usepackage[dvipsnames]{xcolor}
\usepackage{booktabs}
\usepackage{pifont}

\newcommand{\cmark}{\ding{51}}
\newcommand{\xmark}{\ding{55}}

\title{Game-Time: Evaluating Temporal Dynamics in Spoken Language Models}
\name{\shortstack{Kai-Wei Chang\sthanks{Co-first authors}\textsuperscript{1}, 
      En-Pei Hu\footnotemark[1]\textsuperscript{2}, 
      Chun-Yi Kuan\textsuperscript{2}, 
      Wenze Ren\textsuperscript{2}, 
      Wei-Chih Chen\textsuperscript{2}, \\
      Guan-Ting Lin\textsuperscript{2}, 
      Yu Tsao\textsuperscript{3}, 
      Shao-Hua Sun\textsuperscript{2, 4},
      Hung-yi Lee\textsuperscript{2, 4}, 
      James Glass\textsuperscript{1}}}
      
\address{\shortstack{\textsuperscript{1} Massachusetts Institute of Technology, USA \\
         \textsuperscript{2} National Taiwan University, Taiwan \\
         \textsuperscript{3} Academia Sinica, Taiwan} \\
         \textsuperscript{4} NTU Artificial Intelligence Center of Research Excellence (NTU AI-CoRE), Taiwan}
\begin{document}
\ninept
\maketitle
\begin{abstract}
Conversational Spoken Language Models (SLMs) are emerging as a promising paradigm for real-time speech interaction. However, their capacity of temporal dynamics, including the ability to manage timing, tempo and simultaneous speaking, remains a critical and unevaluated challenge for conversational fluency. To address this gap, we introduce the \textbf{Game-Time Benchmark}, a framework to systematically assess these temporal capabilities. Inspired by how humans learn a language through language activities, Game-Time consists of basic instruction-following tasks and advanced tasks with temporal constraints, such as tempo adherence and synchronized responses. Our evaluation of diverse SLM architectures reveals a clear performance disparity: while state-of-the-art models handle basic tasks well, many contemporary systems still struggle with fundamental instruction-following. More critically, nearly all models degrade substantially under temporal constraints, exposing persistent weaknesses in time awareness and full-duplex interaction. The Game-Time Benchmark provides a foundation for guiding future research toward more temporally-aware conversational AI. Demos and datasets are available on our project website\footnote{\url{https://ga642381.github.io/Game-Time}}.
\end{abstract}


\section{Introduction}
\label{sec:intro}

In the pursuit of human-like conversation with machines, the research frontier is moving beyond text-based Large Language Models (LLMs). The next challenge lies in mastering conversational dynamics in real-time speech, which has given rise to the field of conversational Spoken Language Models (SLMs)~\cite{ji2024wavchat, cui2025slm, arora2025landscape, wu2024towards, latif2023sparks, hu25f_interspeech, fu2025vita}. This marks a critical shift from rigid turn-by-turn dialogues to fluid spoken interactions. 
Achieving such dynamics requires SLMs to operate in a \emph{real-time full-duplex} manner~\cite{ma2025language, chen2025minmo}, where models must listen and speak simultaneously while producing seamless responses. This is inherently difficult, demanding synchronous speech generation, continual intent recognition, and precise control over both what to respond and when to respond.
A further challenge lies in modeling the temporal dynamics of spoken interaction, where contemporary systems often fail to capture the fine-grained timing that is essential for advanced conversational fluency. For instance, they struggle to process user speech while planning a coherent reply that aligns with user-specified timing and tempo. This limitation underscores a fundamental deficiency: the lack of \emph{time-awareness}, which is essential for real-world deployment, enabling SLMs to deliver time-critical instructions (e.g., emergency guidance) and to synchronize speech with user activities (e.g., healthcare voice agents).
\begin{figure}[t]
    \centering
    \includegraphics[width=\columnwidth]{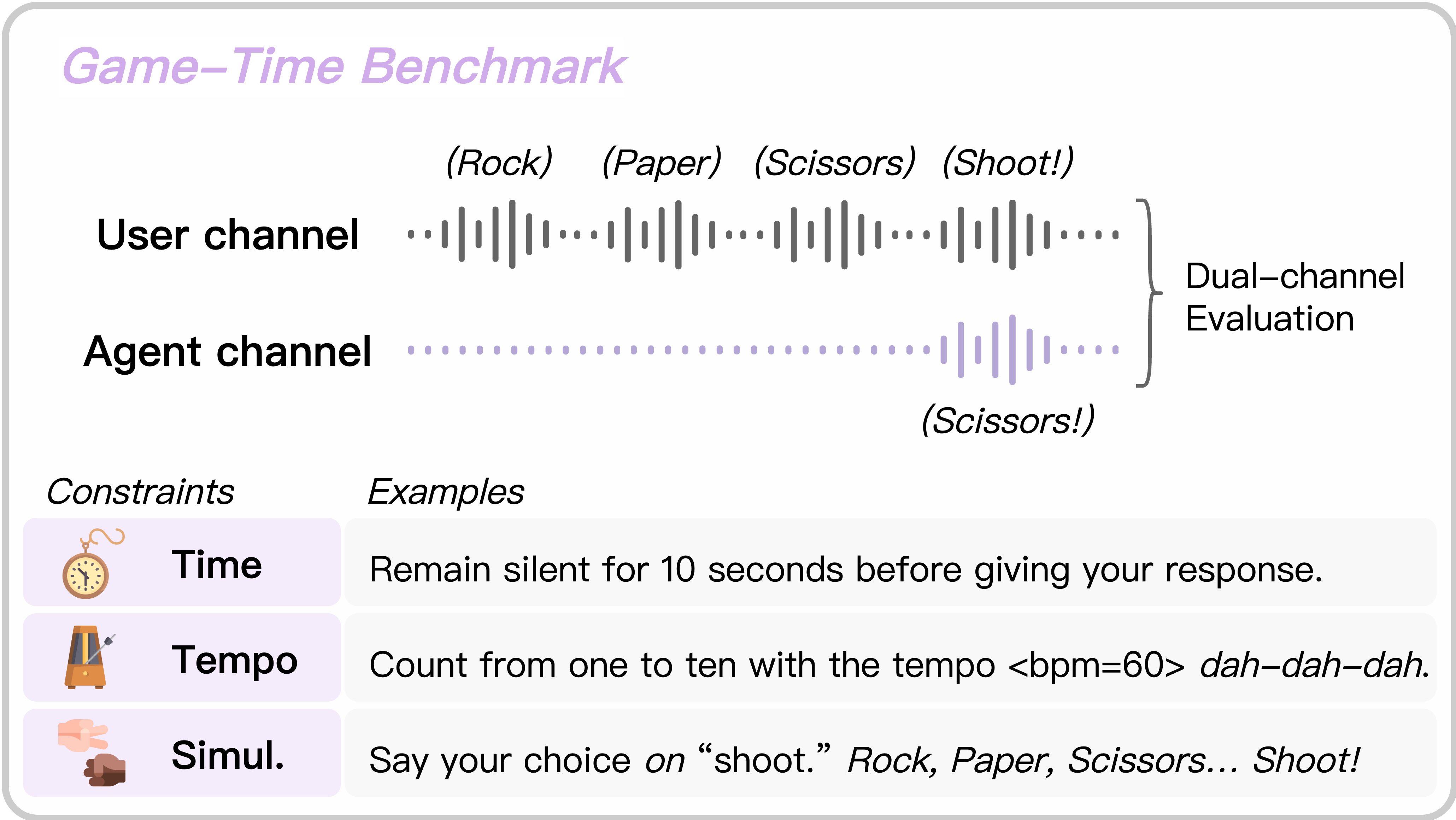}
    \caption{Overview of the \emph{Game-Time Benchmark}, evaluating temporal dynamics in conversational Spoken Language Models (SLMs).}
    \label{fig:overview}
\end{figure}

\begin{table*}[t!]
\centering
\caption{Game-Time Task Families. Basic Task contains fundamental tasks. Advanced Task contains temporal dynamics paired with Basic Task. $\mathcal{N}$: Number of subtasks. *The game of Rock paper scissors is itself an advanced task.}
\vspace{2pt}
\label{tab:gt-overview}
\renewcommand{\arraystretch}{0.93} 
\begin{tabularx}{\textwidth}{@{}c l c p{3.6cm} X@{}}
\toprule
Category & Task Family & $\mathcal{N}$ & Subtask / Paired Basic Tasks & Description \\
\midrule
\midrule
\multirow{6}{*}{Basic}
  & \makebox[1em][c]{1}~-~Sequence       & 3 &Number; Alphabet; Spell & Generate sequential items in order from $n_\text{start}$ to $n_\text{end}$. \\
  & \makebox[1em][c]{2}~-~Repeat         & 2 &Word; Sentence          & Repeat user-provided content $\mathcal{C}$. \\
  & \makebox[1em][c]{3}~-~Compose        & 2 &Word; Scenario          & Compose response including a target word $w$ or fitting a scenario $\mathcal{S}$. \\
  & \makebox[1em][c]{4}~-~Recall         & 3 &Vocabulary; Letter; Rhyme & Name $N$ items that satisfy property $\phi$. \\
  & \makebox[1em][c]{5}~-~Open-Ended     & 2 &Empathy; QA             & Provide helpful and contextually appropriate content. \\
  & \makebox[1em][c]{6}~-~Role-Play      & 2 &Scenario; Persona       & Act within an imagined scenario $\mathcal{S}$ or play a given persona $\mathcal{P}$. \\
\midrule
\multirow{7}{*}{Adv.}
  & \makebox[1em][c]{A}~-~Time-Fast     & 10 &[Multiple Basic Tasks] & Complete the task quickly, within a specified duration $\tau_{\text{fast}}$. \\
  & \makebox[1em][c]{B}~-~Time-Slow     & 7 &[Multiple Basic Tasks] & Perform the task slowly, taking at least the specified duration $\tau_{\text{slow}}$. \\
  & \makebox[1em][c]{C}~-~Time-Silence  & 4 &Repeat; Recall; Open-Ended & Insert a silent interval of $s$ seconds before the response. \\
  & \makebox[1em][c]{D}~-~Tempo-Interval     & 4 &Sequence; Recall & Follow a specified tempo with $\delta$-second space between each word. \\
  & \makebox[1em][c]{E}~-~Tempo-Adhere       & 4 &Sequence; Recall & Adhere to the tempo specified by the user’s spoken example $\mathcal{C}_\text{tempo}$. \\
  & \makebox[1em][c]{F}~-~Simul.-Shadow  & 1 &Repeat & Repeat each word with immediate, word-by-word overlap. \\
  & \makebox[1em][c]{G}~-~Simul.-Cue     & 1 &Rock paper scissors* & Overlap with the user by speaking at a designated timing or cue. \\
\bottomrule
\end{tabularx}
\end{table*}



Existing benchmarks for SLMs focus on content and style generation quality~\cite{yan2025uro, yang2025towards}, mimicking human dialogue behaviors (e.g. back-channeling)~\cite{lin2025full}, and turn-taking~\cite{aroratalking}. However, they lack the focus on temporal dynamics in the conversation. 
To address this critical gap, we introduce the \textbf{\emph{Game-Time Benchmark}}, which is designed to assess the temporal capabilities of SLMs, focusing on their ability to perceive, predict, and produce speech in sync with the user. 

This work is inspired by childhood language acquisition~\cite{bruner1985child}. Children learn to talk through language activities that require not only understanding the meaning of vocabulary, such as counting numbers and naming objects, but also acquiring a sense of timing and tempo in coordination games~\cite{ratner1978games, whitebread2017role}. For example, playing ``rock-paper-scissors'' relies on sharing a common tempo and acting precisely on a specific cue (e.g., “shoot”). This inspiration led us to design the Game-Time Benchmark, which contains two categories of tasks.
The \textbf{Basic Tasks} evaluate an SLM's foundational ability to follow simple instructions, a challenge that can still prove difficult for modern SLMs. The \textbf{Advanced Tasks} build upon this foundation by augmenting the core instructions with temporal constraints. 
Here, the SLM must perform the basic tasks while fulfilling requirements for timing and synchronicity.
The Game-Time Benchmark provides a framework for evaluating whether a model can move beyond mere content generation to acquire the temporal dynamics of conversational fluency. 
It proposes a novel perspective on evaluation, focusing not just on \emph{what to say}, but critically, on \emph{when to say}.

In this paper, we evaluate various SLMs with different design philosophies, including Moshi~\cite{defossez2024moshi}, Unmute~\cite{zeghidour2025streaming}, Freeze-Omni~\cite{wang2025freeze}, Gemini-Live~\cite{google2025geminiLiveAug}, and GPT-realtime~\cite{openai2025gptrealtime}.
Our results show a performance disparity even on Basic Tasks: while state-of-the-art models generally excel, many contemporary SLMs still struggle with fundamental instruction-following. Furthermore, the performance of nearly all models degrades significantly when temporal constraints are introduced. Our findings indicate that models especially struggle with tasks requiring time awareness and real-time full-duplex capability, revealing a critical gap in the capabilities of current systems.
For reproducibility, datasets and results are available on our website.
\section{Related Works}
\subsection{Full-duplex Spoken Language Models}
Recent work has explored how SLMs can move beyond turn-based interaction toward full-duplex conversation~\cite{lin2025full, nguyen2023generative, chenreinforcement, yu2025salmonn, xu2025qwen2, zhang2025omniflatten}, where listening and speaking occur simultaneously. Two main modeling strategies have emerged to achieve full-duplex capability~\cite{arora2025landscape}: 

\textbf{(1) Dual-channel SLMs}~\cite{nguyen2023generative, defossez2024moshi, wang2025ntpp, wu2025aligning} directly process two channels: a listening channel for user speech and a speaking channel for the model’s own output. 
Although this architecture significantly increases modeling complexity, it naturally supports real-time listening and speaking concurrently.

\textbf{(2) Time-multiplexing SLMs}~\cite{zhang2024beyond, wang2025freeze, zeghidour2025streaming} include a state prediction mechanism~\cite{wang2024full} that decides whether to speak or remain silent. The model monitors for turn-taking cues during user speech and switches to speech generation until external interruption occurs.

In the Game-Time benchmark, we evaluate both model designs and commercial voice agent APIs, comparing their ability to manage timing and overlap, and discussing the trade-offs of each design.

\subsection{Benchmarks for Conversational Spoken Language Models}
A number of benchmarks have been proposed to evaluate SLMs. Some
focus on the spoken language understanding and paralinguistics generation~\cite{yan2025uro, ao2024sd, chen2024voicebench, gao2025adubench, liu2025vocalbench, zhang2025wildspeech}. 
Others target interaction naturalness, including \emph{Full-Duplex-Bench}~\cite{lin2025full}, which evaluates SLM's conversational abilities such as interruption handling and overlap management;
\emph{Talking Turns}~\cite{aroratalking}, which focuses on turn-taking dynamics and conversational flow in dialogue;
and \emph{FD-Bench}~\cite{peng25b_interspeech}, which assesses the SLM's responsiveness and robustness under noisy scenarios. More recently,
\emph{Beyond Words}~\cite{liao2025beyond} proposes a multimodal framework to determine when a model should speak by modeling turn-taking and backchanneling in audiovisual conversations.
While these benchmarks characterize important conversational behaviors, they do not directly evaluate whether models can comprehend and satisfy explicit temporal constraints.


\section{Game-Time Benchmark}
We introduce the \emph{Game-Time Benchmark} to evaluate SLMs on their understanding of \emph{time}, \emph{tempo}, and timely \emph{simultaneously speaking}. In this section, we define the task families, describe how the benchmark is constructed, and outline the evaluation protocol.
\vspace{-5pt}
\subsection{Task Families}
Inspired by how humans learn a language with language activities and games,
the Game-Time benchmark comprises two categories: Basic Tasks, testing fundamental speech capabilities, and Advanced Tasks, which paired temporal constraints with suitable Basic Tasks to assess the model's time-awareness. The full taxonomy of families, subtasks, and temporal requirements is summarized in Table~\ref{tab:gt-overview}.

\textbf{Game-Time Basic Tasks: }  
The 6 Basic Task families reflect fundamental capabilities of spoken interaction.  
They are inspired by the kinds of activities through which humans first practice language: reciting ordered sequences (\emph{Sequence}), repeating spoken content (\emph{Repeat}), composing sentences that meet specific criteria (\emph{Compose}), recalling items from memory (\emph{Recall}), responding helpfully and openly in conversation (\emph{Open-ended}), and navigating hypothetical scenarios or adopting specific personas (\emph{Role-play}). 
Together, they capture a spectrum from structured behaviors (e.g., counting) to open-ended and social behaviors (e.g., \emph{Empathy} and \emph{Role-Play}). 

\textbf{Game-Time Advanced Tasks: }  
Advanced tasks introduce constraints that move beyond \emph{what} to say and focus on \emph{when} to say it.  
Here, the philosophy is to test temporal and interactive fluency — skills that come naturally to humans but remain underexplored in SLMs.
\emph{Time} tasks examine whether models can modulate overall duration of speaking time, which require coordinating not only the speaking rate but also the content; \emph{Tempo} tasks probe their ability to sustain rhythmic consistency or synchronize with an external beat; \emph{SimulSpeak} tasks challenge them to overlap with the user’s speech, listening and synchronizing with the user in real time.  
These constraints are abstractions of conversational dynamics such as timing, tempo and coordination, which are central to human conversation. 



\subsection{Task Formalization}
We formalize our tasks within an Instruction-Following (IF) framework~\cite{zhou2023instruction, pyatkin2025generalizing}. Each IF instance is specified by a base task $t$ and a set of constraints $\mathcal{C}$. Performing an IF task requires the model to \emph{perform a base task $t$ while satisfying all constraints in $\mathcal{C}$}. Each constraint is a predicate over variables that are typically numeric or symbolic. 

\textbf{Example:}
Consider the user instruction, ``Please count from one to ten in $10$ seconds.'' Here, the base task is \emph{sequential generation} $t_\text{seq}$. Also, this instruction implies two constraints: (i) a range constraint $c_\text{range}$with variables $(n_\text{start}, n_\text{end})=(1,10)$ requiring the spoken sequence to be $1,2,\dots,10$, and (ii) a duration constraint $c_\text{dur}$with variable $\tau_\text{fast}=10\text{s}$ requiring the task performing time to be less than 10 seconds. For this instance, the model should perform the base task $t_\text{seq}$ while satisfying $\mathcal{C}=\{c_\text{range}, c_\text{dur}\}$. Building upon this formalization, we can systematically generate the dataset by creating natural language templates, instantiating them with diverse variables, and synthesizing the resulting instructions into speech.

\begin{figure}[t]
    \centering
    \includegraphics[width=0.9\columnwidth]{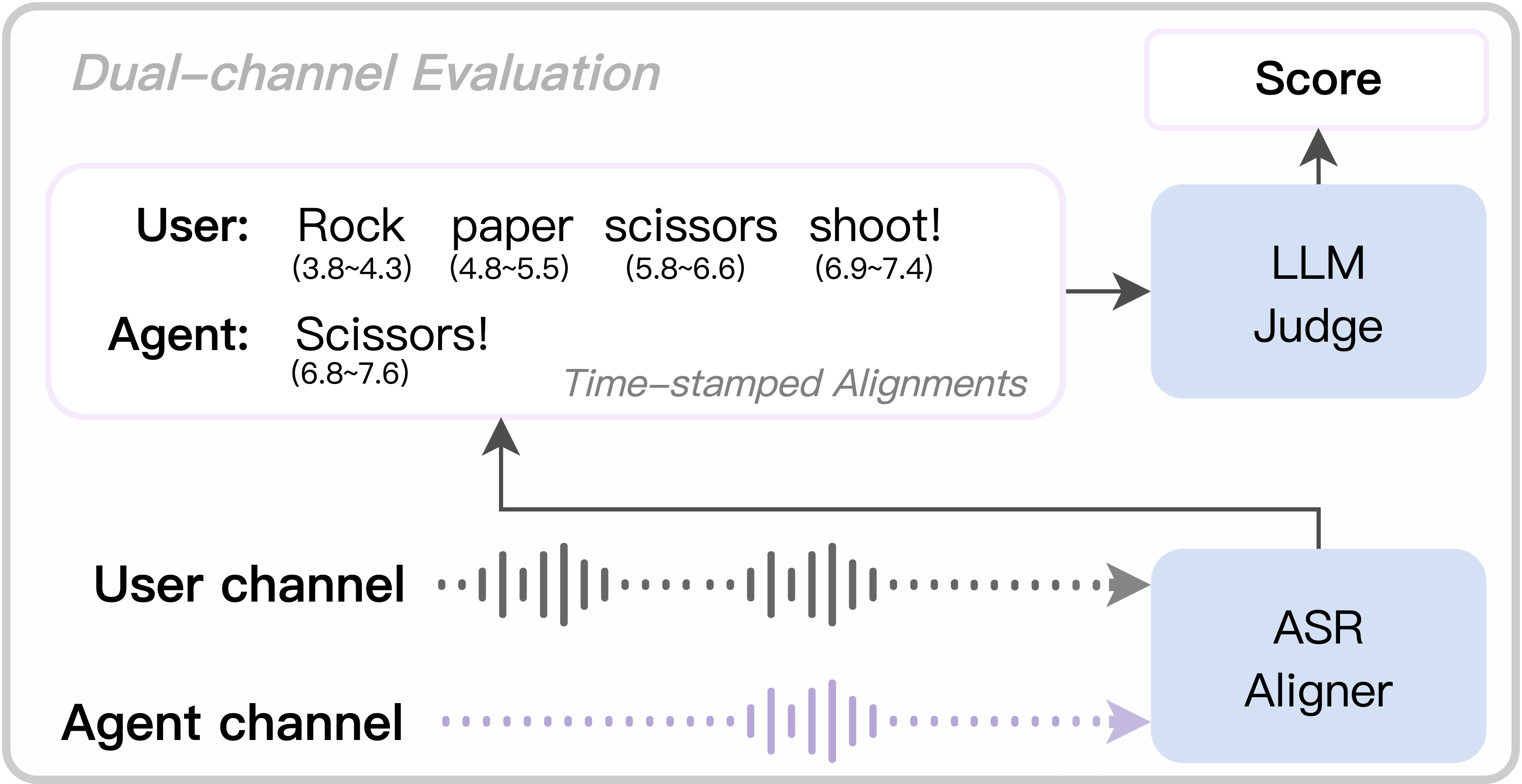}
    \caption{Dual-channel Evaluation with LLM-as-a-judge.}
    \label{fig:evaluation}
\end{figure}

\subsection{Dataset Construction Pipeline}
The Game-Time benchmark dataset is constructed through a four-stage pipeline designed to generate a diverse and high-quality set of spoken instructions. \textbf{(i) Seed Instruction Creation}: We begin by manually writing a set of seed instructions for Basic Tasks, defining the base task and corresponding variables.
\textbf{(ii) Linguistic Diversification}: An LLM paraphrases seed instructions to create a variety of linguistic templates. We then populate these templates by varying the defined variables, resulting in a large and diverse set of instruction texts for the Basic Tasks. The Advanced Tasks are derived by augmenting a subset of Basic Tasks with temporal constraints $C$. This controlled approach ensures that corresponding basic and Advanced Tasks share the same underlying base task $t$, allowing us to focus on the performance variation when imposing constraints.
\textbf{(iii) Speech Synthesis}: These text-based instructions are synthesized into audio using a TTS system with multiple voices to ensure vocal diversity~\footnote{We primarily use CosyVoice~\cite{du2024cosyvoice} for speech synthesis. For tasks requiring precise tempo control, the audio was edited manually. Google TTS was used for the ``Rock, Paper, Scissors'' task, as we found it produced higher-quality output in this case.}. 
\textbf{(iv) Quality Control}: Finally, we use an ASR model to transcribe the synthesized audio. Instructions whose transcriptions do not closely match the original text are filtered out. This automated check is supplemented by manual listening verification on a majority of the samples to ensure high perceptual quality.

In the end we have a total of 1,475 test instances: 700 samples for the Basic tasks (14 subtasks, 50 samples each) and 775 samples for the Advanced tasks (31 subtasks, 25 samples each).

\subsection{Dual-channel Evaluation}
Our evaluation protocol, illustrated in Figure~\ref{fig:evaluation}, leverages an LLM to score model performance. For each dialogue, we first transcribe the dual-channel audio (user and model) to obtain time-aligned text. This transcription is then provided to an LLM judge, which assesses the model's performance on the criteria of instruction following
\footnote{For ``Open-Ended'' Basic Tasks, which lack an explicit instruction, the LLM judge evaluates ``response appropriateness'' instead.}.

We also explored alternative methods: using an \emph{audio-LLM-as-a-judge}~\cite{chiang2025audio} and employing rule-based automatic metrics. We found the audio-LLM approach is also effective but more costly and less aligned with human's evaluation (discussed in Sec.\ref{subsec:human-eval}).
Meanwhile, rule-based metrics are often too rigid for the interpretive nature of spoken dialogue. For instance, a rigid script would penalize a model for including a natural conversational preamble (e.g., ``Okay, I'll start now...'') in a time-constrained task. In contrast, an LLM can perform reasoning to recognize and evaluate the core speech embedded in the whole dialogue and give a reasonable evaluation\footnote{While we could have designed instructions to be unequivocally precise, doing so would result in unnatural conversations. Our approach prioritizes evaluating a model's ability to interpret natural instructions and time-related cues, rather than its ability to follow overly constrained commands.}. 

Overall, we find this text-based LLM judge to be a unified, simple, yet effective method. To validate this approach, we conduct subjective human evaluations and confirm that its assessments align with human preferences. Furthermore, this method can effectively evaluate other behaviors such as turn-taking. Due to space limitations, these additional results are available on our project website.

\section{Experimental Setup}
\label{sec:models}

\begin{table}[t]
\centering
\caption{Comparison of SLMs in Game-Time Benchmark.}
\vspace{2pt}
\label{tab:full_duplex_models}
\resizebox{0.9\columnwidth}{!}{%
\begin{tabularx}{\columnwidth}{llcc}
\toprule
\textbf{Model} & \textbf{Full-Duplex Method} & \textbf{Open} & \textbf{Frozen LLM}\\
\midrule
Freeze-Omni   & Time-Multiplexing      & \color{OliveGreen} \cmark & \color{OliveGreen} \cmark\\
Unmute        & Time-Multiplexing      & \color{OliveGreen} \cmark & \color{OliveGreen} \cmark\\
Moshi         & Dual Channel           & \color{OliveGreen} \cmark & \color{Red} \xmark\\
Gemini-Live   & --                     & \color{Red} \xmark & -     \\
GPT-realtime  & --                     & \color{Red} \xmark & -\\
\midrule
SSML-LLM      & Non-causal Completion  & -- & \color{OliveGreen} \cmark \\
\bottomrule
\end{tabularx}
}
\end{table}

\begin{figure*}[t]
    \centering
    \includegraphics[width=\textwidth]{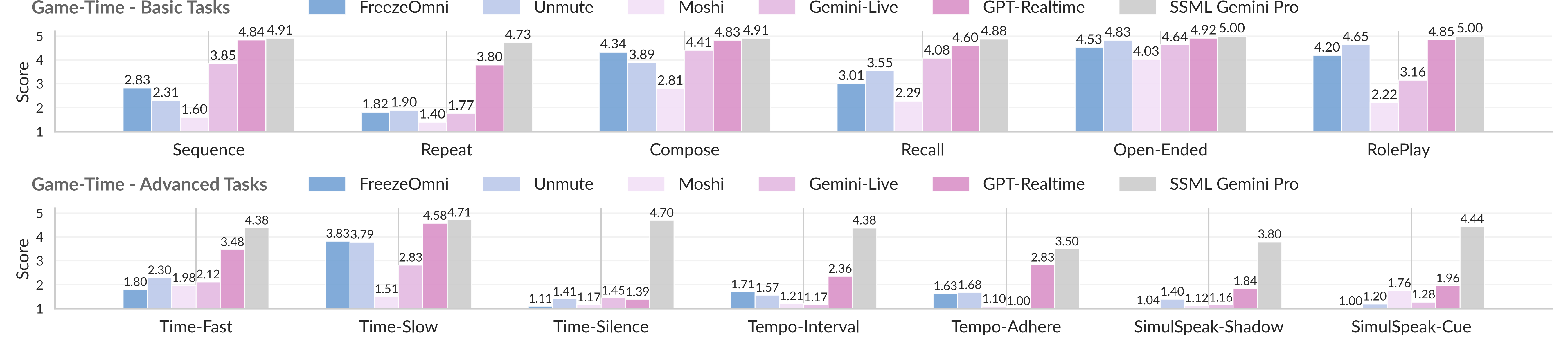}
    \vspace{-20pt}
    \caption{Game-Time benchmark scores evaluated with LLM-as-a-judge. \textbf{Top}: results on Basic Tasks. \textbf{Bottom}: results on Advanced Tasks.}
    \label{fig:llm}
\end{figure*}

\begin{figure*}[t]
    \centering
    \includegraphics[width=0.98\textwidth]{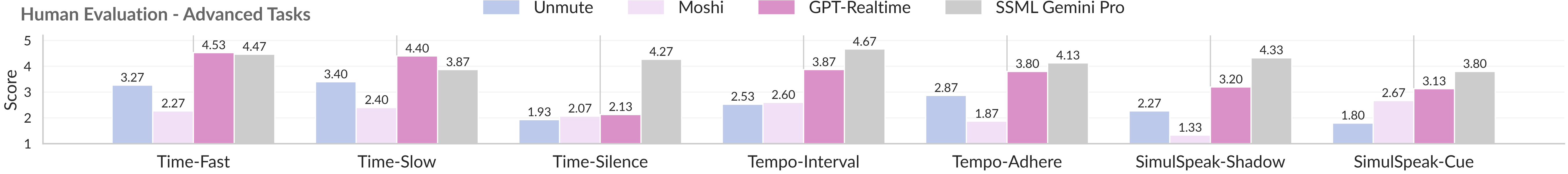}
    \vspace{-10pt}
    \caption{Human evaluation on Game-Time Advanced Tasks.}
    \label{fig:adv-human}
\end{figure*}

We evaluate various SLMs on the Game-Time Benchmark with different full-duplex strategies (see Table~\ref{tab:full_duplex_models}). This includes \textbf{Time-Multiplexing} models (Freeze-Omni~\cite{wang2025freeze}, Unmute~\cite{zeghidour2025streaming}) which use a modular pipeline of a streaming encoder, a frozen LLM, and a streaming decoder; and a \textbf{Dual-channel} model (Moshi~\cite{defossez2024moshi}) where a fine-tuned LLM directly processes and generates speech. We also include \textbf{commercial voice agents}: Gemini-Live\footnote{Gemini-Live~\cite{google2025geminiLiveAug}. API model name: \emph{gemini-live-2.5-flash-preview}} and GPT-realtime\footnote{GPT-realtime~\cite{openai2025gptrealtime}. API model name: \emph{gpt-realtime}}

\textbf{Oracle System:}
We introduce \emph{SSML-LLM} as our benchmark's oracle topline. This is a non-streaming and non-causal system that operates with future knowledge. It receives the full word-level alignments of the user's utterance as input, and an LLM then generates a dialogue counterpart with timing that is precisely controlled and synchronized with the user's speech via Speech Synthesis Markup Language (SSML). This SSML is then synthesized into audio by a TTS system. For example, after processing the user's speech “Rock paper scissors shoot!”, it generates \texttt{<ssml><break time="6.9s"/>scissors!</ssml>} to make the response ``Scissors!'' overlap with user's ``Shoot!''. Although not feasible in real-time, SSML-LLM helps calibrate our LLM judge and human evaluations by providing a theoretical performance ceiling. We use Gemini 2.5 Pro~\cite{comanici2025gemini} with reasoning and Google TTS.


\textbf{Dual-channel Evaluation}: 
Our LLM-as-a-judge framework requires time-stamped data for its analysis. To obtain word-level alignments, we use the Whisper-medium model. Based on a preliminary study, we selected Gemini 2.5 Pro as the LLM judge, which we found superior to other open-source and commercial models at processing these time-stamped transcripts, leveraging its strong reasoning capabilities for evaluating complex temporal behaviors.

\section{Results}
\subsection{Main Results}
\hspace*{1.6em}\textbf{Basic Tasks:} As shown in Fig.~\ref{fig:llm} (Top), the oracle topline consistently achieves the best performance across all tasks. GPT-realtime shows strong performance on most Basic Tasks, and it is worth noting that in \emph{Repeat}, it is the only model that delivers reasonable performance.
On the other hand, we observe that time-multiplexing models (Freeze-Omni and Unmute), which rely on a frozen LLM, generally outperform the dual-channel model (Moshi). This suggests that fine-tuning a text LLM to model speech signals remains challenging in spoken conversation scenarios.
Overall, although Basic Tasks can be handled by the most advanced model (GPT-realtime), there is still room for improvement in modern academic models.

\textbf{Advanced Tasks:} As shown in Fig.~\ref{fig:llm} (Bottom), introducing temporal constraints results in a substantial drop in performance. Among the Advanced Tasks, models perform comparatively better on \emph{Time-Fast} and \emph{Time-Slow}, but fail on \emph{Time-Silence}, suggesting that they can adjust their speaking rate in response to user instructions but still fail to grasp precise temporal requirements. 
Similarly, adhering to tempos (\emph{Tempo}) and synchronizing speech with users (\emph{SimulSpeak}) remains difficult for modern SLMs, even for SOTA commercial voice agents such as GPT-realtime. This performance disparity suggests that current SLMs do not possess time-awareness, highlighting the need to focus on this capability in future research.

\subsection{Human Evaluation}
\label{subsec:human-eval}
\begin{table}[t]
\centering
\caption{Correlation between human judge with LLM and ALLM judge, both using Gemini 2.5 Pro.}
\vspace{2pt}
\renewcommand{\arraystretch}{0.85} 
\begin{tabular}{lcc}
\toprule
  & Spearman’s $\rho$ & Pearson’s $r$ \\
\midrule
Human~-~LLM & 0.677 & 0.675 \\
Human~-~ALLM & 0.643 & 0.625 \\
\bottomrule
\end{tabular}
\label{tab:llm_human_corr}
\end{table}


Fig.~\ref{fig:adv-human} shows the result of the human evaluation. For each task, there are 20 samples, with each evaluated by three human judges via Prolific. We observe a similar trend in the performance of SLMs as in LLM dual-channel evaluation. 
Table~\ref{tab:llm_human_corr} presents the correlation between LLM-as-a-judge scores and human evaluations for Advanced Tasks. We also list the Audio LLM judge score for reference. Across 4 models $\times$ 35 data scores, we observe a reasonably high correlation (Spearman's $\rho=0.677$, Pearson's $r=0.675$) for our dual-channel evaluation method. 
These results suggest that the LLM-as-a-judge is reliable and well aligned with human evaluation. We also find that for tasks requiring precise measurements like maintaining a ten-second silence, the LLM may be more objective than humans, as it can leverage time-stamped alignment data for evaluation.

\section{Conclusion}

This paper introduced the Game-Time Benchmark to address a critical gap in the evaluation of the temporal dynamics of conversational Spoken Language Models (SLMs). 
We evaluated various SLMs with a series of tasks testing temporal capabilities of timing, tempo, and simultaneous speaking.
Our results reveal a clear gap, with some models able to handle basic instructions, but nearly all failing once temporal constraints are introduced.
This widespread inability to manage precise timing or synchronize with a user reveals a persistent lack of time-awareness in current SLMs, even in the most advanced systems. Our dual-channel evaluation utilizing an LLM for reasoning was shown to be a reliable method, offering a unified and scalable way to measure these complex behaviors. We hope the Game-Time Benchmark will motivate the community to build the next generation of diverse and time-aware SLMs.




\section{Acknowledgments}
We are grateful to Yi-Cheng Lin and Cheng-Han Chiang for their valuable discussions on evaluation methods, and to Shih-Yun Shan Kuan for assistance with commercial API usage.

This work was supported by the Ministry of Education (MOE) of Taiwan under the project Taiwan Centers of Excellence in Artificial Intelligence, through the NTU Artificial Intelligence Center of Research Excellence (NTU AI-CoRE).

Kai-Wei Chang was supported in part by the National Science and Technology Council (NSTC), Taiwan, under Grant No. 114-2917-I-564-024.

\bibliographystyle{IEEEbib}
\bibliography{refs}
\end{document}